\documentclass[aps,prd,twocolumn,groupedaddress,nofootinbib]{revtex4}

\usepackage{graphicx}
\usepackage{amsmath,amssymb,amsfonts}
\usepackage[colorlinks,citecolor=blue,linkcolor=blue,urlcolor=blue]{hyperref}
\usepackage[applemac]{inputenc}

\usepackage{physics} % for \bracket \matrixel \Re \Im
\usepackage[percent]{overpic} %writing over pictures

\usepackage[english]{babel}

\usepackage[autostyle, english = american]{csquotes}
\MakeOuterQuote{"}

\newcommand{\be}{\begin{equation}}
\newcommand{\ee}{\end{equation}}

\begin{document}

\title{Can we travel to the past? Irreversible physics along closed timelike curves}

\author{Carlo Rovelli}
\email{rovelli.carlo@gmail.com}
\affiliation{Aix-Marseille University, Universit\'e de Toulon, CNRS, CPT, 13288 Marseille, France.\\
Perimeter Institute, 31 Caroline Street North, Waterloo, Ontario N2L 2Y5, Canada.\\
The Rotman Institute of Philosophy, 1151 Richmond St.~N
London, Ontario N6A 5B7, Canada.
}

\date{\small\today}

\begin{abstract}
\noindent
The Einstein equations allow solutions containing closed timelike curves.  These have generated much puzzlement and suspicion that they could imply paradoxes.  I show that puzzlement and paradoxes disappears if we  discuss carefully the physics of the irreversible phenomena in the context of these solutions. 
\end{abstract}

\maketitle 

\section{Introduction}

Traveling to the distant \emph{future} is possible in principle: it suffices to travel fast enough, or to spend little  time sufficiently near to a black hole horizon.   But can we travel to the \emph{past}? 

General relativity appears to suggest that this might be possible in principle, because it allows solutions with closed timelike curves \cite{Godel1949,Thorne1992}.   But the idea of traveling to the past has always raised puzzlement.  It is commonly said that traveling to the past would generate logical paradoxes, such as the possibility of killing our own parents before we were born. 

Here I argue that this and similar paradoxes disappear if we examine with care the full thermodynamical and statistical physics along the closed timelike curves. The paradoxes disappear, without any need of resorting to quantum physics, as suggested for instance in \cite{Deutsch1991,Ringbauer2014}.   (See also \cite{Lewis2000} for a general philosophical discussion of the problem, and \cite{Nikolic2006} where the same basic idea illustrated here is also presented.) 

\section{A clock around a closed timelike curve}

Consider a clock coupled to the gravitational field and whose worldline follows a closed timelike curve. Let  $ds^2=g_{ab}(x)dx^a dx^b$ be a solution of the Einstein equations (with the clock) containing a closed timelike curve $\gamma$.  Here $a,b=0,1,2,3$  and I assume  signature [+,-,-,-] for simplicity of notation. The  proper time along the entire timelike curve is 
$
S=\int_\gamma ds.
$
Writing the curve explicitly in coordinates, $\gamma:  \tau\mapsto \gamma^a(\tau), \tau\in[0,1]$, the proper time grows along the curve grows as 
\be
s(\tau)= \int_0^\tau d\tau' \sqrt{g_{ab}(\gamma(\tau'))\, \dot \gamma^a(\tau')\dot \gamma^b(\tau')} 
\ee 
where $\dot\gamma^a\!=\!d\gamma^a/d\tau$.  A closed timelike curve satisfies $\gamma(1)\!=\!\gamma(0)$ and its normal is everywhere timelike, that is 
\be
g_{ab}\dot \gamma^a\dot \gamma^b>0. 
\ee
Notice that since $s(0)\!=\!0$ but $s(1)\!=\!S$, $ds$ is not an exact one-form on $\gamma$: that is, it is not the differential of a continuous function $s$ on $\gamma$. 

A clock is a mechanical device including for instance a harmonic oscillator beating proper time.  Let $\theta$ be the angular variable describing the position of the hand of the clock, and $\omega=2\pi/T$ its frequency.   The position of the hand of the clock along the path is given by
\be
\theta(\tau)=\theta(0)+\omega \, s(\tau),\ \rm{modulo}\  2\pi
\ee 
and the number of oscillations since the start is the (real) number 
\be
n(\tau)=\frac{\omega}{2\pi} \, s(\tau)=\frac{s(\tau)}{T} .
\ee

If the two physical variables $g_{ab}(x)$ and $\theta(\tau)$ satisfy the equations of motion it follows necessarily that $\theta(0)=\theta(1)$.  That is, the physical equations impose that the hand of the clock "comes back at the end of the loop" to the same position as it was at the start of the loop.  In other words, the total number of oscillations around the line, $N=n(1)=S/T$, must be integer.  If $N$ is not an integer, the equations of motion are not satisfied, and the theory tells us that this is an impossible state of affairs. 

This observation might seem to trivially eliminate any paradox but it appears to be strange: intuitively it seems that if I follow a closed timelike curve I get back to the initial spacetime point but I can have a memory of having being around, therefore there should be something different in my final configuration, with respect to my departing configurations, after going round the loop.  This, after all, is what intuitively mean by "traveling back to the past". 

To investigate this, let us focus simply on the total amount $S=N\,T$ of proper time along the closed timelike line $\gamma$. This is measured by the integer $N$.  Any clock capable of measuring $N$ would indeed get back to the initial spacetime location with a record of having been around. 

Can the clock keep track of the number $N$ of its oscillations?  Physical clocks do that regularly: they not only beat the period, but  have also a device that records the number of oscillations. 

However, now comes the main point of this paper: any mechanical clock that counts oscillations dissipates energy.   That is, any clock is ultimately thermodynamical. The escapement of a pendulum clock for instance cannot work without  friction.   This fact has been emphasized  for instance by Eddington \cite{Eddington1928}. It is also discussed in a wonderful lecture by Feynman \cite{Feynmana}.  This is the key observation of this paper.  Let us see what it implies, disregarding, for the moment, the statistical mechanics underpinning thermodynamics (to which I return later).

To have a clock counting oscillations, we need dissipation, hence entropy to grow.   Let $S(\tau)$ be a measure of local entropy along the closed loop $\gamma$. For the clock to work all along $\gamma$, registering the number of its oscillations, we need $dS(\tau)/d\tau>0$ everywhere going around the loop.  

But since $\gamma$ is a loop, $S(\tau)$ cannot grow monotonically as we go round.   Therefore it is impossible for a clock to count its own oscillations along a closed timelike curve.   There is no physical way for a clock to count its oscillations along a closed timelike curve. This conclusion has far reaching consequences, discussed below. 

\section{Travel to the past is thermodynamically impossible}

The above conclusion is in fact far more general than clock oscillation counting.   For instance, if we want to travel to the past and arrive to the past keeping our memory of events happened in the future, we need some device (like our brain) capable of memory.   But memory is an irreversible phenomenon (we remember the past not the future) and, like all irreversible phenomena, is based solely on the only fundamental irreversible law: the second principle of thermodynamics $dS(\tau)/d\tau\ge 0$.  Along a closed timelike loop $\gamma$ the only possibility of having $dS/d\tau\ge 0$ everywhere is having $dS/d\tau= 0$. But this means that all the processes around $\gamma$ are reversible, and therefore there can be no memory. 
 
The phenomenology that we commonly associate to the forward passage of time (memory, decisions, cumulative counting of the oscillations of a periodic device....) depends \emph{entirely} on the second principle of thermodynamics.  This implies that the future direction is determined by the derivative of a state function, the entropy.  Since no function can uniformly increase around a circle, we can never "travel to the past" in the sense of arriving to the past having memory of the future, having counted the oscillations of our clock, being in a position of acting different that what we did, or similar. These are all  thermodynamical phenomena (that require irreversibility), and  thermodynamics does not permit travel to the past in this sense.  

General relativity allows closed timelike curves, and this is not incompatible with anything.   But along these curves entropy cannot grow monotonically and therefore they cannot be uniformly future-oriented in the sense of any  phenomenon that distinguish the past from the future.  The thermodynamical arrow of time, which is the one that determines the phenomena we associate to the forward passage of time, cannot loop back onto itself, because it is a gradient. 

In other words, the proper time measured by a \emph{reversible} periodic device is mathematically described by a closed one-form $ds$ which can be well defined along a closed path.  But the time that distinguishes the past from the future is a thermodynamical quantity $dS$ that is the differential of a state function, and therefore is exact, and therefore cannot grow uniformly along a circle. 

What happens in the case in which along a closed timelike curve the entropy increases and decreases in $\tau$?  Consider the simplest possibility where $dS/d\tau > 0$ for $\tau\in \gamma_+=[0,\hat\tau]$ and $dS/d\tau < 0$ 
for $\tau\in \gamma_-=[\hat\tau, 2\pi]$.   Then the above discussion immediately clarifies the physics of this solution: for everything that concerns irreversible phenomena such as memory, decisions and keeping track of the past, the effective direction of time is towards increasing $\tau$ in $\gamma_+$ and towards decreasing $\tau$ in $\gamma_-$.  See Figure 1.  All paradoxes disappear.  

\begin{figure}

\includegraphics[width=5cm]{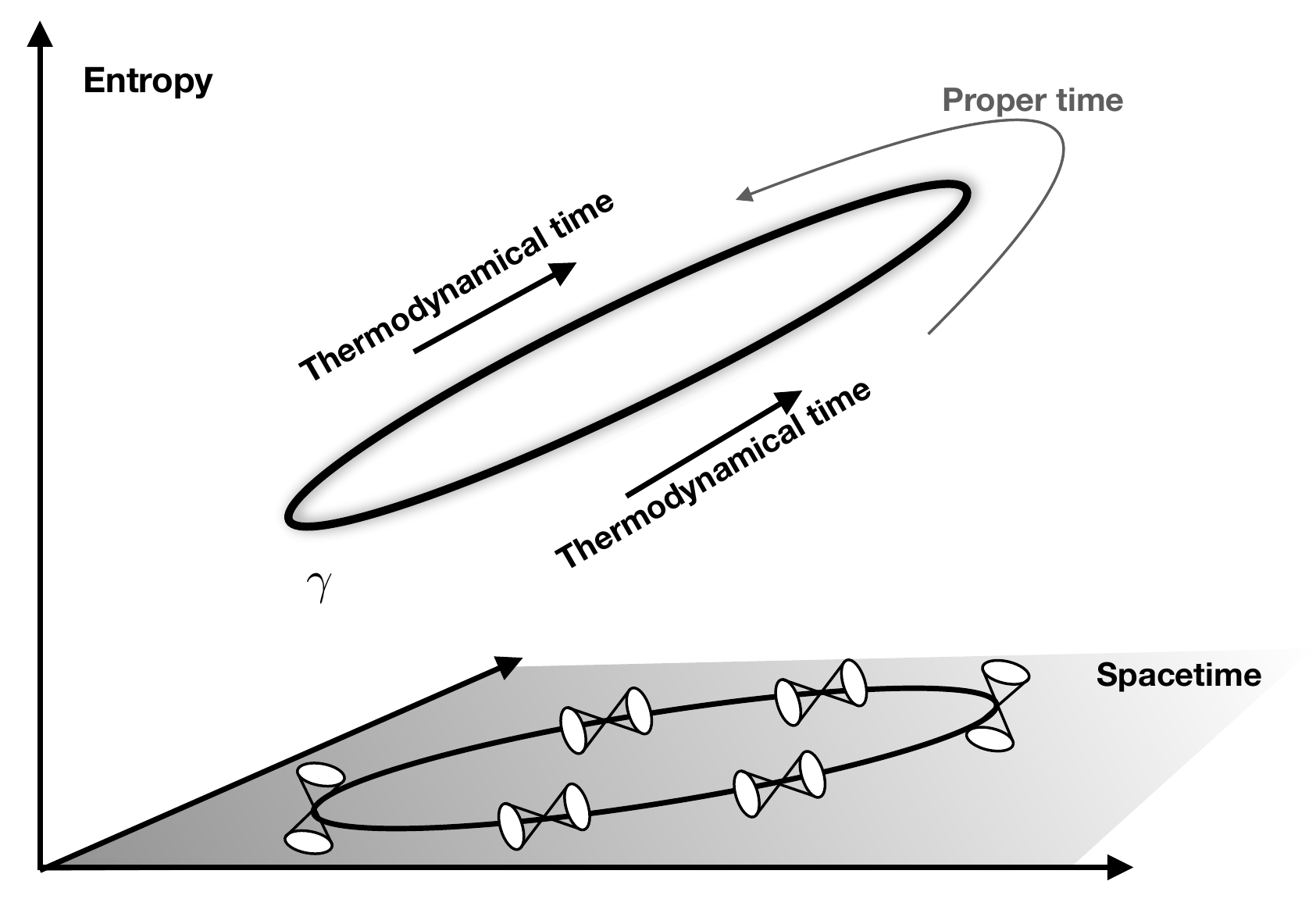}
\caption{The direction of increasing proper time around a closed timelike curve does not agree with the future direction of the thermodynamical arrow of time.}
\end{figure}

The above discussion clarifies  the physics of closed timelike curves and their relation with our intuition that  "time moves only forward".   Our intuition is based on the fact that all phenomena that we see as characteristic of "forward moving time" are thermodynamical phenomena where entropy grows.  Since entropy cannot grow constantly along a closed loop, we cannot travel forward in time and return to a previous space time location in this sense, {\em even if spacetime admits closed timelike curves}. 

\section{The statistical mechanics picture}

Thermodynamics is nothing but mechanics restricted to a relatively small number of "macroscopic" ("coarse-granined") variables, and under the condition that the entropy defined by this coarse-graining was low in some region, denoted "past".   Any thermodynamical statement can therefore be in principle translated into a statistical mechanical statement, conditional to past low entropy.   What is the statistical account of the thermodynamical impossibility of traveling back in time derived above? 

The key, as is always the case in the relation between thermodynamics and statistical mechanics, is a careful distinction between the microscopic account and the macroscopic account.  Let us  disentangle the two. 

In the microscopic account, the notion of irreversibility makes no sense.   Irreversibility, indeed, is the property of certain \emph{macroscopic} histories to be far more likely in one time direction than in the opposite time direction, under the condition of past low-entropy. (A glass that breaks and a broken glass that repairs spontaneously are  macroscopic histories.)  All \emph{microscopic} histories, on the other hand, are equally improbable; in fact, they are equally improbable whether considered ahead or back in time.  The restriction to histories that are low-entropy in the past breaks time reversal invariance.  (The number of microscopic histories where a glass breaks --back or forward in time-- is small, but the initial low-entropy state happens to include precisely a class that does so in one direction and not in the other.) Hence there is no irreversible phenomenon at the microscopic level. At this level, we cannot talk about phenomena like memory, and there is no sense in which the arrival point of a closed timelike loop can be distinguished by its starting point.   The universe is what it is at every spacetime location and (in classical mechanics) the future and the past are equally fully determined by the present. 

In the macroscopic account, on the other hand, we lump together large numbers of microscopic histories. By doing so, we loose the determinism of classical mechanics.  A macroscopic configuration can evolve into different futures and may come from different pasts. This opens the space for talking about "decision", "memory", and similar irreversible phenomena, which describe peculiar aspects of macro-histories.  

If the entropy determined by the coarse graining is low in some spacetime region, this generates a characteristic  phenomenology which determines the difference between the past and the future.   

The application of all this to closed timelike curves is simple: macroscopic histories where time appear uniformly oriented all along the curve are, simply: very improbable.    We  might figuratively say that traveling to the future is possible, while traveling to the past is extremely improbable. 

\section{Conclusion}

The closed timelike curves of general relativity generate no paradoxes.  But they do not allow us to travel to past in the thermodynamical sense --- for instance being in the past having memory of the future. 

 The key to the solution of the apparent paradoxes posed by the closed timelike curves in general relativity is to distinguish different meanings of the expression "time".  Most confusion about time originates from mixing different uses of the word "time" \cite{Rovelli2018}.  The time of mechanics, which is not oriented and is simply determined by the oscillations of a clock can unproblematically come back to itself along a general relativistic closed timelike curve.   But the thermodynamical time, namely the oriented time of thermodynamical phenomena, including our experiential time, cannot. \\

\centerline{***}

I thank Steven Savitt for a discussion that has lead to this paper. 
This work was partially supported by the FQXi  Grant  FQXi-RFP-1818.

%\bibliographystyle{utcaps}
%\bibliography{/Users/carlorovelli/Documents/library}
%
%\end{document}

\providecommand{\href}[2]{#2}\begingroup\raggedright\endgroup

\end{document}